\documentclass{mem}
\usepackage{natbib}
\usepackage[varg]{txfonts}
\usepackage{balance}
\usepackage{graphicx}
\usepackage[a4paper,breaklinks,dvips]{hyperref}
\idline{75}{282}
\begin{document}
\def\teff{$T\rm_{eff }$}
\def\logg{$\log g$}
\def\moh{[M/H]}
\def\kms{$\mathrm {km s}^{-1}$}
\def\ximic{$\xi_{\rm mic}$}
\def\ximac{$\xi_{\rm mac}$}
\def\vhydr{$v_{\rm \,hydro}$}
\newcommand{\ximicM}[1]{$\xi_{\rm mic}^{\rm \,M#1}$}

\title{
Micro- and macroturbulence predictions from CO5BOLD 3D stellar atmospheres
}

   \subtitle{}

\author{
M. Steffen\inst{1,3}
\and 
E.~Caffau\inst{2,3}
\and
H.-G.~Ludwig\inst{2,3}
       }

\offprints{M. Steffen}

\institute{
Leibniz Institut f\"ur Astrophysik Potsdam, 
An der Sternwarte 16, D-14482 Potsdam, Germany, \email{msteffen@aip.de}
\and
Zentrum f\"ur Astronomie der Universit\"at Heidelberg, Landessternwarte,
K\"onigstuhl 12, D-69117 Heidelberg, Germany
\and
GEPI, Observatoire de Paris, CNRS, Universit\'{e} Paris Diderot, 
Place Jules Janssen, 92190 Meudon, France
}

\authorrunning{Steffen et al.}

\titlerunning{Micro- and macroturbulence from 3D models}

\abstract{
We present an overview of the current status of our efforts to derive 
the microturbulence and macroturbulence parameters (\ximic\ and \ximac)
from the CIFIST grid of CO5BOLD 3D model atmospheres as a function of the 
basic stellar parameters \teff, \logg, and \moh. The latest results for the 
Sun and Procyon show that the derived microturbulence parameter 
depends significantly on the numerical resolution of the underlying 3D 
simulation, confirming that `low-resolution' models tend to underestimate
the true value of \ximic. Extending the investigation to
$12$ further simulations with different \teff, \logg, and \moh, we obtain
a first impression of the predicted trend of \ximic\ over the 
Hertzsprung-Russell diagram: in agreement with empirical evidence,
microturbulence increases towards higher effective temperature and lower
gravity. The metallicity dependence of \ximic\ must be interpreted with care,
since it also reflects the deviation between the 1D and 3D photospheric
temperature stratifications that increases systematically towards lower \moh.
 
}

\maketitle{}

\section{Introduction}
In the context of classical spectrum analysis based on 1D model atmospheres, 
the auxiliary parameters micro- and macroturbulence (\ximic\ and \ximac) play 
an important role. One of the great advantages of 3D hydrodynamical model
atmospheres is their physically consistent thermal structure and velocity
field, properly taking into account the effect of convective flows, overshoot, 
and waves. Based on first principles without free parameters, 3D stellar
atmosphere models can be used to derive \ximic\ and \ximac\ from their
hydrodynamical velocity fields. Even if these classical turbulence
parameters are not relevant when using 3D model atmospheres for spectroscopic 
work, their knowledge can be useful. On the one hand, such results can be 
compared with empirical data obtained from classical 1D studies to check the 
validity of the 3D models. On the other hand, the predictions of the 3D models 
can be used to fix \ximic\ and \ximac\ in 1D spectroscopic work when 
an empirical determination of these parameters is not possible, due to a lack of
microturbulence-sensitive lines (e.g.\ in low-resolution spectra of very 
metal-poor stars), or due to the ambiguity between macroturbulence and 
rotational line broadening.

In the following, we review the different methods we have used so far
to derive the parameters $\xi_{\rm mic}$ (and subsequently $\xi_{\rm mac}$) 
from  our 3D model atmospheres, and give an update of the relevant results.
Our investigations are mainly focused on the Sun 
(\teff=5780~K, \logg=4.44, [M/H]=0) and Procyon (\teff=6500~K, \logg=4.0, 
[M/H]=0), where we have also studied the influence of the spatial resolution
of the 3D numerical simulations on the derived turbulence parameters.
Based on $12$ further simulations with different \teff, \logg, and \moh, we
finally report first preliminary results of an investigation in progress
aiming at the prediction of the variation of \ximic\ over the 
Hertzsprung-Russell diagram (HRD). All 3D model atmospheres used for this
investigation are taken from the CIFIST\footnote{extended with respect to 
the 2009 grid by additional models representing prominent real stars} 
3D model atmosphere grid \citep{Ludwig+al09} computed with 
CO5BOLD\footnote{http://www.astro.uu.se/$\sim$bf/co5bold\_main.html} 
\citep{Freytag+al12}. 

\section{Derivation of \ximic\ and \ximac\ from 3D~model atmospheres}
The different methods we have used to derive the turbulence parameters 
\ximic\ and \ximac\ from 3D hydrodynamical model atmospheres have been
introduced by \citet{Steffen+al09}. We give a brief summary of the
relevant points here.

\subsection{Microturbulence}
For simplicity, and in line with common practice, all methods assume 
that the small-scale photospheric velocity field may be 
described by an isotropic Gaussian probability distribution of 
the line-of-sight velocity, $P(v)$\,$\sim$\,$\exp(-v^2/\xi_{\rm mic}^2)$,
characterized by a single, depth-independent parameter, \ximic.
Basically, the value of \ximic\ is inferred from its effect on the 
equivalent width ($W$) of synthetic spectral lines, not from its influence
on the shape of the line profiles.
 
{\bf Method~1} (M1) 
relies only on the 3D model, and yields a value of \ximic\ for 
any individual spectral line, thus allowing to map \ximic\ as a function
of line strength $W$ and/or excitation potential $E_i$ for arbitrary ions.
In principle, this method can also be employed to map a possible 
depth-dependence of \ximic\ in the hydrodynamical model by selecting lines 
forming at different atmospheric heights.

Given the spectral line parameters, the line profile is computed from the 
3D model with different velocity fields: (i) using the original 
3D hydrodynamical velocity field, and (ii) replacing the 3D hydrodynamical 
velocity field by an isotropic, depth-independent microturbulence, like in 
classical 1D spectrum synthesis, but retaining the full 3D thermodynamic 
structure. The microturbulence associated with the considered spectral line,
\ximicM{1}, is defined by the requirement: 
$W_{\rm 3D}($\vhydr$) = W_{\rm 3D}(\,$\ximicM{1}$)$, 
where $W_{\rm 3D}($\vhydr$)$ and $W_{\rm 3D}(\,$\ximicM{1}$)$ 
are the equivalent widths obtained in steps (i) and (ii), respectively.
The procedure works well even for lines as weak as  $W \approx 10$~m\AA.

Method~1 is considered the most powerful and flexible procedure to extract the 
`true' microturbulence parameter from a 3D numerical convection simulation,
because it measures only the effect of the non-thermal velocity field on
the line formation, excluding the additional influence of thermodynamic
fluctuations. However, it is also the computationally most expensive method,
and cannot be applied to the analysis of observed stellar spectra.

{\bf Method~2a/b} (M2a/b)
In this method, \ximic\ is not derived for a single spectral line, but for 
a given ion with fixed excitation potential $E_i$. The concept relies on
a \emph{set of fictitious spectral lines} generated from a 
\emph{curve-of-growth}, i.e. all lines share the same atomic parameters 
except for the oscillator strength ($\log gf$), which controls the line 
strength. The microturbulence parameter is then defined with the help of a 
suitable 1D reference model atmosphere. 

Given any set of fictitious spectral lines, we first compute for each 
line $i$ the equivalent width from the 3D model, $W_{\rm 3D}\,(\log g f_i)$.
Next we compute for each of the lines a 2-dimensional curve-of-growth
from the adopted 1D reference model,
$W_{\rm 1D}\,(\log g f_i + \Delta \log \epsilon, \,\xi_{\rm mic})$, 
where $\Delta \log \epsilon$ and \ximic\ are the independent variables.
For fixed \ximic\, the 1D-3D abundance difference 
$\Delta \log \epsilon_i(\,\xi_{\rm mic})$ is now defined by the condition 
$W_{\rm 3D}\,(\log g f_i) = 
W_{\rm 1D}\,(\log g f_i + \Delta \log \epsilon_i, \,\xi_{\rm mic})$.
For each line, $\Delta \log \epsilon_i(\,\xi_{\rm mic})$ thus indicates 
the difference between the abundance derived from the 1D model by fitting the
equivalent of the 3D line, and the true abundance assumed in the 3D 
spectrum synthesis. In general, this difference varies from line to line in
a systematic way.

In M2a, we compute $\beta($\ximic$)$, the slope of the linear regression to 
the data set $\{W_{{\rm 3D},\, i},\; \Delta \log \epsilon_i(\,$\ximic$)\}$,
and define \ximicM{2a} by the usual condition that the abundance (correction) 
must not show any systematic trend with line strength,  
i.e.\ $\beta($\ximicM{2a}$)=0$. Alternatively, 
in M2b, we consider $\sigma($\ximic$)$, the standard deviation of the data
set $\{\Delta \log \epsilon_i\}$. In this case, the microturbulence is 
defined by the requirement that it minimizes the standard deviation, 
$\sigma(\,$\ximicM{2b}$)$\,=\,$\min$. M2a and M2b would give exactly the same
microturbulence if $\partial\,(\Delta \log \epsilon_i)\,/\,\partial \xi_{\rm mic}
= -\alpha\,W_{{\rm 3D},\, i}$, with $\alpha$ independent of line $i$, which is
often a good approximation.

{\bf Method~3a/b} (M3a/b)
is a generalization of Method~2a/b. Instead of utilizing a 
set of fictitious lines lying on a single curve-of-growth,
M3 works with a sample of \emph{real} spectral lines, selected to
cover a sufficient range in line strength, with an unavoidable variation 
in excitation potential and wavelength. Again, the 1D-3D abundance correction
$\Delta \log \epsilon_i($\ximic$)$ is computed from the preferred 1D model,
and the value of \ximic\ is adjusted to minimize the difference 
in $\Delta \log \epsilon_i$ between weak and strong lines
(M3a: zero slope for \ximicM{3a}), or to minimize the overall 
dispersion of $\Delta \log \epsilon_i$ (M3b: minimum standard deviation
for \ximicM{3b}). Method~3a corresponds to the classical definition of 
$\xi_{\rm mic}$, applied to synthetic spectra. 

Note that all 3 methods have the advantage that errors in $\log g f$
cancel out. Only Method~3 can also be applied to observed stellar
spectra; in this case the \emph{relative} precision of the $\log g f$ 
values \emph{is} crucial. Obviously, the results depend on the selected 
spectral lines, and, for Methods~2 and 3, on the choice of the 1D reference
 model atmosphere.

{\bf Method~4} (M4)
In principle, it should be possible to derive \ximic\ (and also
\ximac) directly from evaluating the 3D hydrodynamical
velocity field without resorting to synthetic spectral lines. 
A possible concept for such a procedure is described in the Appendix.

\subsection{Macroturbulence}
\label{s:macro}
We assume that the large-scale photospheric velocity field may be 
characterized by a single macroturbulence parameter, \ximac, 
and that macroturbulence can be described by an isotropic 
Gaussian probability distribution of the line-of-sight velocity, 
$P(v)$\,$\sim$\,$\exp(-v^2/\xi_{\rm mac}^2)$. As for the microturbulence, 
the value of \ximac\ is determined from the comparison of 1D and 3D synthetic 
spectral line profiles. While microturbulence affects the process of line
formation and changes both the width and the strength of a spectral line,
the effect of macroturbulence can be described by a subsequent convolution 
of the emergent line profile with the Gaussian macroturbulence profile, a
simple operation that preserves the equivalent width of the spectral line.

The value of \ximac\ can thus only be determined after the microturbulence
parameter \ximic\ has been derived by any of the methods described above. 
Irrespective of the method used to derive \ximic, the macroturbulence 
parameter is determined on a line-by-line basis. Specifically, the
macroturbulence associated with any given spectral line is defined by 
minimizing the mean square difference ($\chi^2$) between the original 3D line
profile $F_{{\rm 3D},\,i}(\lambda)$, computed with the full hydrodynamical 
velocity field, and the line profile obtained from the reference model 
atmosphere\footnote{In case of M1, this is the 3D
model with \vhydr\ replaced by \ximic, otherwise the preferred 1D model.},
$F_{{\rm ref},\,i}(\lambda, \Delta \log \epsilon_i,\, $\ximic$,\,$\ximac$)$. 
The latter line profile is computed with $\Delta \log \epsilon_i$ and 
\ximic\ fixed to the values derived in the previous step, 
ensuring that the two line profiles being compared have the same 
equivalent width. The remaining free parameter \ximac\ is adjusted such that
$\chi^2$ is minimized. We use the IDL procedure MPFIT 
\citep{Markwardt2009} to find the solution 
$\xi^{\,\ast}_{\rm mac}$ that gives the best fit to the original 3D profile.

\begin{table*}
\caption{3D model atmospheres used in this study. Columns (7) and (8)
refer to the number of opacity bins and the turbulent viscosity used in
the hydrodynamical simulations. Column (9) gives the type of external 
reference atmosphere used for the microturbulence determination with 
methods M2 and M3.}
\label{tab:3dmodels}
\begin{center}
\begin{tabular}{lcccccccc}
\noalign{\smallskip}\hline\noalign{\smallskip}
3D Model & \teff & \logg & [M/H] & grid cells & Volume   & opacity & 
$\nu_{\rm turb}$ & 1D   \\
      & [K]   & [cgs] &       &            & [Mm$^3$] & bins    & 
               & ref. \\
\noalign{\smallskip}\hline\noalign{\smallskip}
Sun: \\
sun59std  & $5774$ & $4.44$ & $0.0 $ & $140^2\times 150$ & 
$5.6^2\times 2.3$ & $12$ & high & HM  \\
sun59x16  & $5764$ & $4.44$ & $0.0 $ & $400^2\times 300$ & 
$5.6^2\times 2.3$ & $12$ & high  & HM  \\
\noalign{\smallskip}\hline\noalign{\smallskip}
Procyon: \\
t65g40mm00std  & $6484$ & $4.00$ & $0.0 $ & $140^2\times 150$ & 
$29.0^2\times 28.9$ & $5$ & high  & LHD05  \\
t65g40mm00x8a  & $6474$ & $4.00$ & $0.0 $ & $280^2\times 300$ & 
$29.0^2\times 28.9$ & $5$ & high  & LHD05  \\
t65g40mm00x8b  & $6473$ & $4.00$ & $0.0 $ & $280^2\times 300$ & 
$29.0^2\times 28.9$ & $5$ & low   & LHD05  \\
\noalign{\smallskip}\hline\noalign{\smallskip}
Main sequence: \\
t45g45mm00n01  & $4509$ & $4.50$ & $0.0 $ & $140^2\times 141$ & 
$4.8^2\times 2.0$ & $5$ & high  & LHD05  \\
t50g45mm00n04  & $4982$ & $4.50$ & $0.0 $ & $140^2\times 141$ & 
$4.9^2\times 2.5$ & $5$ & high  & LHD05  \\
t55g45mm00n01  & $5488$ & $4.50$ & $0.0 $ & $140^2\times 150$ & 
$5.9^2\times 3.5$ & $5$ & high  & LHD05  \\
t59g45mm00n01  & $5865$ & $4.50$ & $0.0 $ & $140^2\times 150$ & 
$6.0^2\times 3.8$ & $5$ & high  & LHD05  \\
t63g45mm00n01  & $6233$ & $4.50$ & $0.0 $ & $140^2\times 150$ & 
$7.0^2\times 4.0$ & $5$ & high  & LHD05  \\
t65g45mm00n01  & $6456$ & $4.50$ & $0.0 $ & $140^2\times 150$ & 
$8.4^2\times 4.0$ & $5$ & high  & LHD05  \\
\noalign{\smallskip}\hline\noalign{\smallskip}
Subgiants: \\
t46g32mm00n01  & $4582$ & $3.20$ & $0.0 $ & $200^2\times 140$ & 
$110^2\times 35.2$ & $5$ & high  & LHD05  \\
t50g35mm00n01  & $4923$ & $3.50$ & $0.0 $ & $140^2\times 150$ & 
$59.7^2\times 29.9$ & $5$ & high  & LHD05  \\
t55g35mm00n01  & $5432$ & $3.50$ & $0.0 $ & $140^2\times 150$ & 
$49.0^2\times 35.3$ & $5$ & high  & LHD05  \\
t59g35mm00n01  & $5884$ & $3.50$ & $0.0 $ & $140^2\times 150$ & 
$89.0^2\times 38.2$ & $5$ & high  & LHD05  \\
\noalign{\smallskip}\hline\noalign{\smallskip}
Giants: \\
t45g25mm00n01  & $4477$ & $2.50$ & $0.0 $ & $140^2\times 150$ & 
$851^2\times 292$ & $5$ & high  & LHD05  \\
t50g25mm00n01  & $4968$ & $2.50$ & $0.0 $ & $160^2\times 200$ & 
$573^2\times 243$ & $5$ & high  & LHD10  \\
\noalign{\smallskip}\hline\noalign{\smallskip}
Metal-poor sun: \\
t57g44mm20n03  & $5734$ & $4.44$ & $-2.0 $ & $140^2\times 150$ & 
$5.5^2\times 2.3$ & $12$ & high  & LHD10  \\
\noalign{\smallskip}\hline\noalign{\smallskip}
Leo star: \\
t59g40mm40n02  & $5850$ & $4.00$ & $-4.0 $ & $200^2\times 200$ & 
$26.0^2\times 12.5$ & $11$ & high  & LHD05  \\
\noalign{\smallskip}\hline\noalign{\smallskip}
\end{tabular}
\end{center}
Notes: HM: Holweger-M\"uller empirical model atmosphere \citep{hmsunmod};
LHD05/10: 1D~mixing-length model ($\alpha_{\rm MLT}=0.5/1.0$) with 
same stellar parameters and opacity scheme as 3D model. 
\end{table*}
\begin{figure*}[htb]
\begin{center}
\mbox{\includegraphics[bb=40 28 570 370,width=0.75\textwidth,clip=true]
{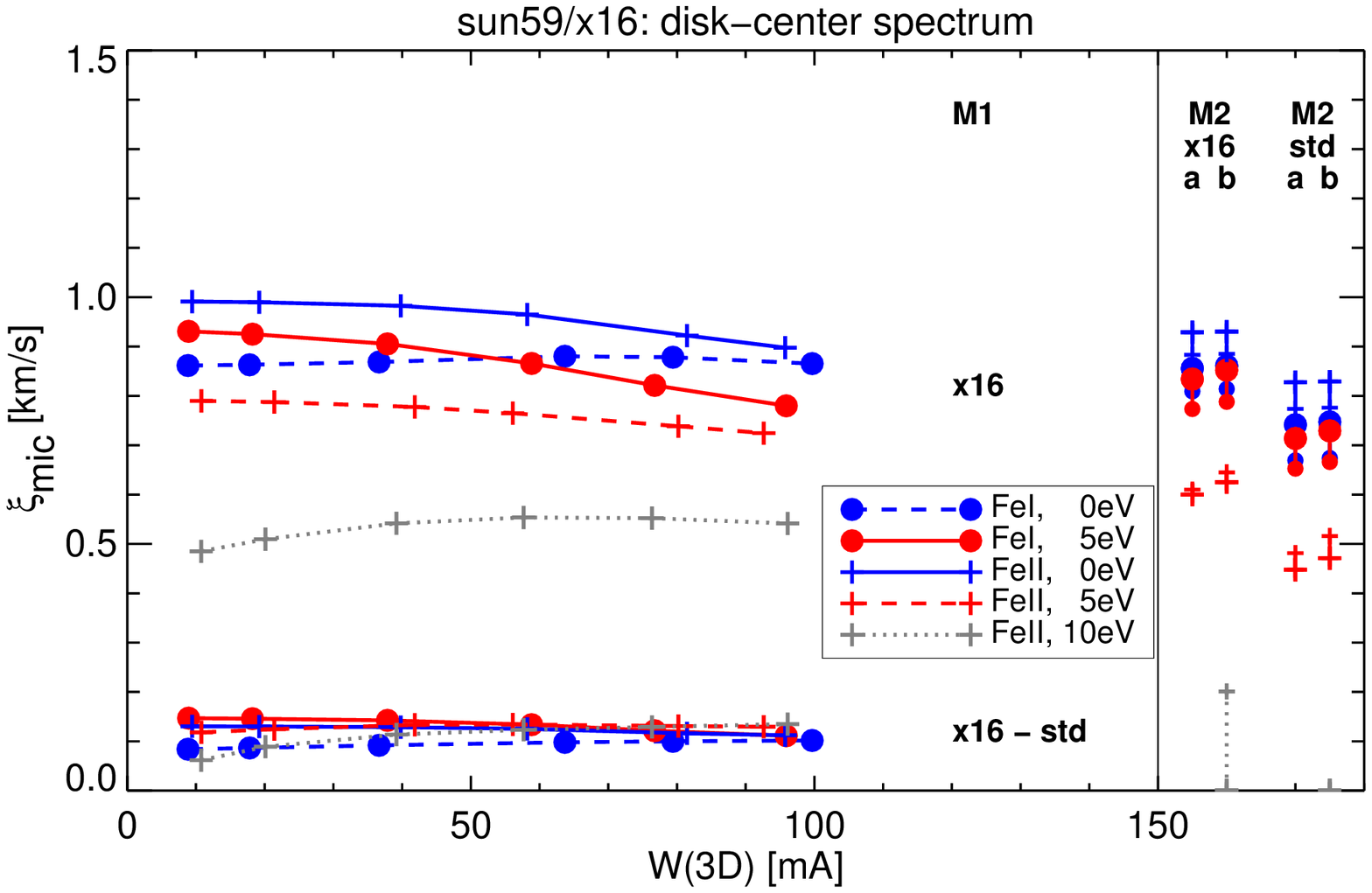}}
\mbox{\includegraphics[bb=40 28 570 370,width=0.75\textwidth,clip=true]
{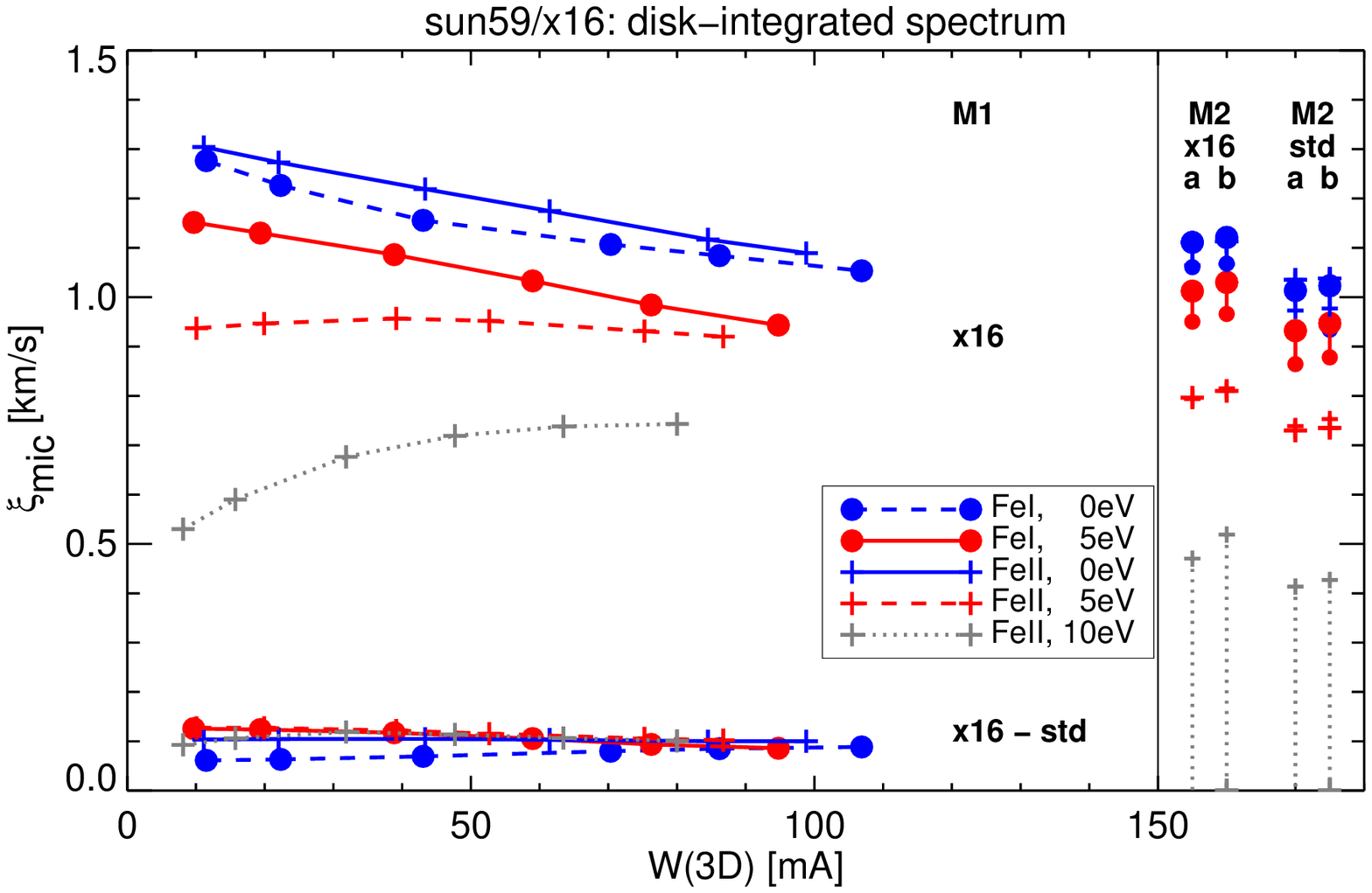}}
\end{center}
\caption{\footnotesize
Determination of the microturbulence parameter \ximic\ for two different
solar models (sun59std and sun59x16) according to methods M1 and M2, using
$5$ sets of $6$ fictitious iron lines each (see legend) at $\lambda\,550$~nm. 
The left part of each panel shows the results of M1, where \ximic(sun59x16) is
derived from individual lines (upper set of curves). The difference
\ximic(sun59x16) - \ximic(sun59std) is indicated by the lower set of curves. The
right part of each panel displays the results of M2, where symbols indicate
the \ximic\ values derived from sets of lines with constant excitation
potential. For each model, \ximic\ has been evaluated with M2a and M2b 
(sun59x16: left columns, sun59std: right columns), using two different 1D 
reference atmospheres ($\langle$3D$\rangle$: smaller symbols, HM: larger 
symbols). Upper and lower panel refer to the disk-center and full-disk
spectrum, respectively.
}
\label{f:sun1}
\end{figure*}
\begin{figure*}[htb]
\begin{center}
\mbox{\includegraphics[bb=40 28 570 370,width=0.75\textwidth,clip=true]
{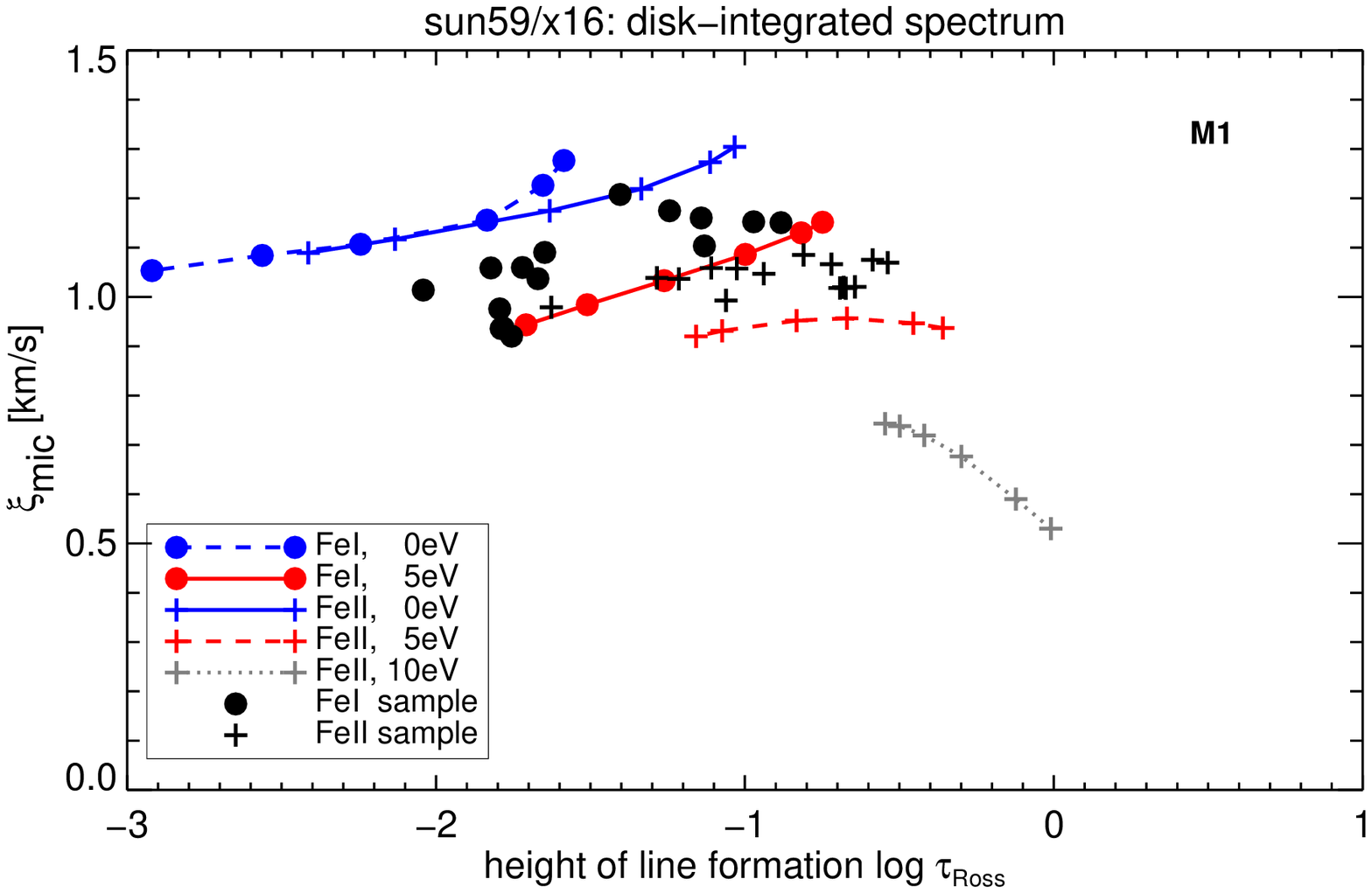}}
\end{center}
\caption{\footnotesize
Microturbulence parameter \ximic\ derived from high-resolution solar model 
sun59x16 according to method M1. The results are the same as shown in
in Figs.\,\ref{f:sun1} and \ref{f:sun3} (flux spectra, lower panels), but 
are here plotted over the line's height of formation. The latter is defined 
as the Rosseland optical depth that divides the line forming region in two 
equal parts, each contributing half of the line's equivalent width.  
}
\label{f:sun2}
\end{figure*}

\section{Results for the Sun and Procyon}
\subsection{\ximic\ for the Sun}
Figure \ref{f:sun1} shows the determination of the 
microturbulence parameter from two different 3D model atmospheres of the
Sun (see Tab.\,\ref{tab:3dmodels}), according to methods M1 and M2.
We considered $5$ sets of fictitious iron lines: \ion{Fe}{i} 
with $E_i$=$0$ and $5$~eV, \ion{Fe}{ii} with $E_i$=$0$, $5$, and $10$~eV. The
 $6$ lines of each set lie on a curve-of-growth with constant $E_i$, while 
$\log gf$ is varied to control the line strength such that it falls in the
range $5$\,m\AA\ $ \la W_{\rm 3D} \la 100$\,m\AA. We assume that all lines have 
the same wavelength, $\lambda = 550$~nm.

\subsubsection{Method 1}
Obviously, the derived value of $\xi_{\rm mic}$ depends on the type and
strength of the considered spectral line. M1 clearly reveals that
high-excitation lines tend to `feel' a lower microturbulence than 
low-excitation lines. At first sight, this is an unexpected result, because
the high-excitation lines tend to form in the deeper photosphere where the 
velocity amplitudes are larger. At the same time, however, the vertical 
extent of the line forming region becomes narrower with increasing excitation
potential, and thus the velocity variation probed by the spectral line
decreases. Apparently, the latter effect dominates and leads to a reduced
microturbulence broadening.

The dependence of \ximic\ on equivalent width $W$ is non-trivial. 
The results obtained from the evaluation of the disk-center spectrum
indicate a rather weak dependence on line strength (Fig.\,\ref{f:sun1},
upper panel), while the disk-integrated spectrum exhibits a more pronounced
\ximic($W$) dependence, the slope of which is a function  of excitation
potential (Fig.\,\ref{f:sun1}, lower panel).

A very general result is that, for a given set of lines, the microturbulence
derived from the disk-integrated (flux) spectra is systematically higher
than that obtained from the disk-center (intensity) spectra, in agreement with 
observational evidence \citep[e.g.][]{Holweger+al78}. However, the ratio
between \ximic(flux) and \ximic(intensity) depends again on the line
properties, ranging from $\approx 1.5$ for weak low-excitation \ion{Fe}{i} 
lines to $\la 1.1$ for weak high-excitation \ion{Fe}{ii} lines.

Finally, we have to point out that the predicted microturbulence values 
depend slightly on the numerical resolution of the 3D model atmospheres used 
for the calculation of the synthetic line profiles.
Increasing the spatial resolution by a factor of $2$ in the vertical
direction, and by a factor $\sqrt{8}$ in each of the horizontal directions
(decreasing $\Delta x$\,=$\,\Delta y$ from $40$ to $14$~km, and $\Delta z$ 
from $15$ to $7.5$~km), leads to an increase of \ximic\ by roughly $0.1$~km/s 
for all type of lines. It remains unclear which numerical resolutions is 
needed to obtain fully converged microturbulence results. A minimum
requirement is, of course, that the spacing of the hydrodynamical grid
must be much smaller than the vertical extent of the line forming region.

\begin{figure*}[htb]
\begin{center}
\mbox{\includegraphics[bb=40 28 570 370,width=0.75\textwidth,clip=true]
{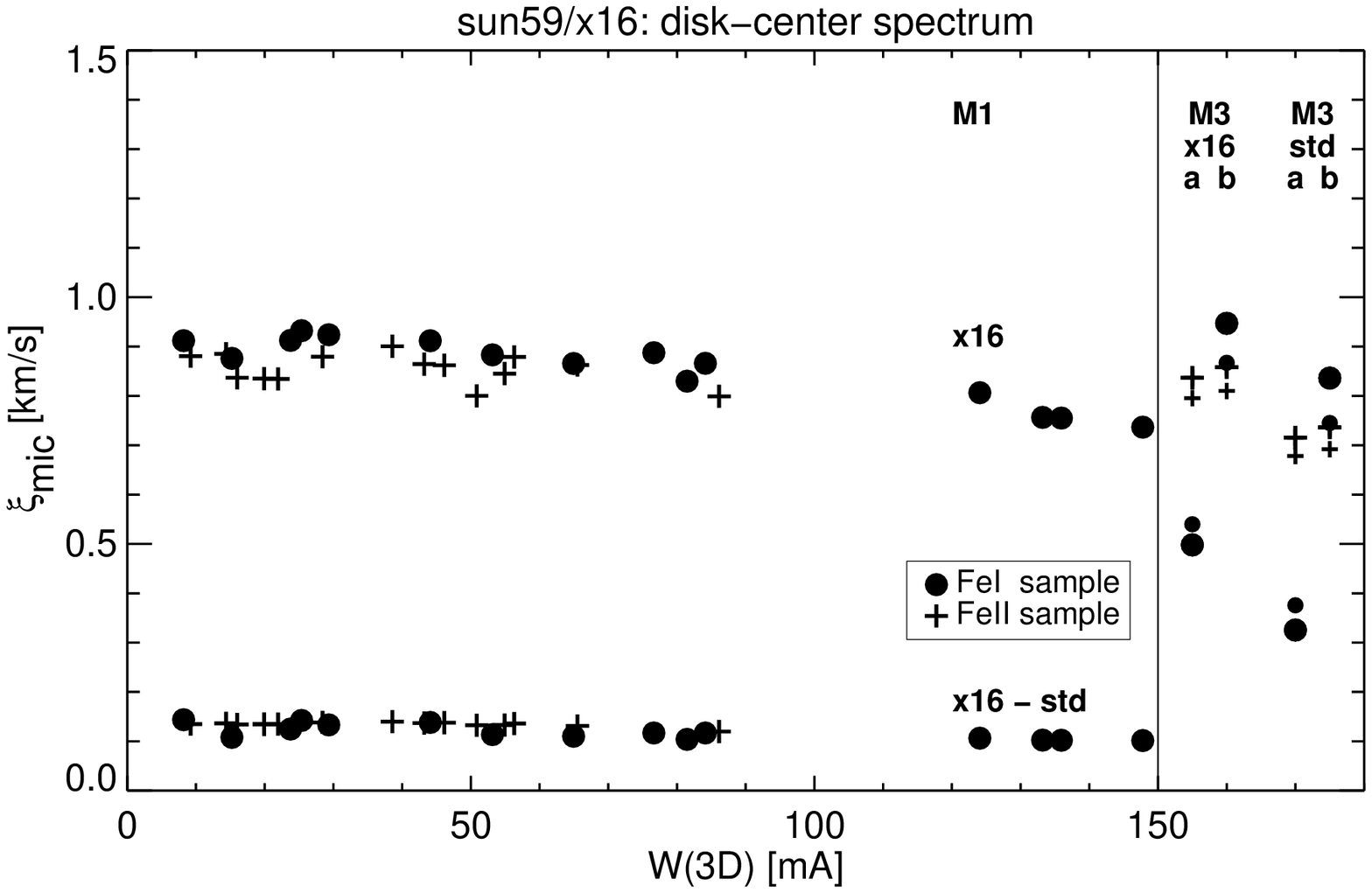}}
\mbox{\includegraphics[bb=40 28 570 370,width=0.75\textwidth,clip=true]
{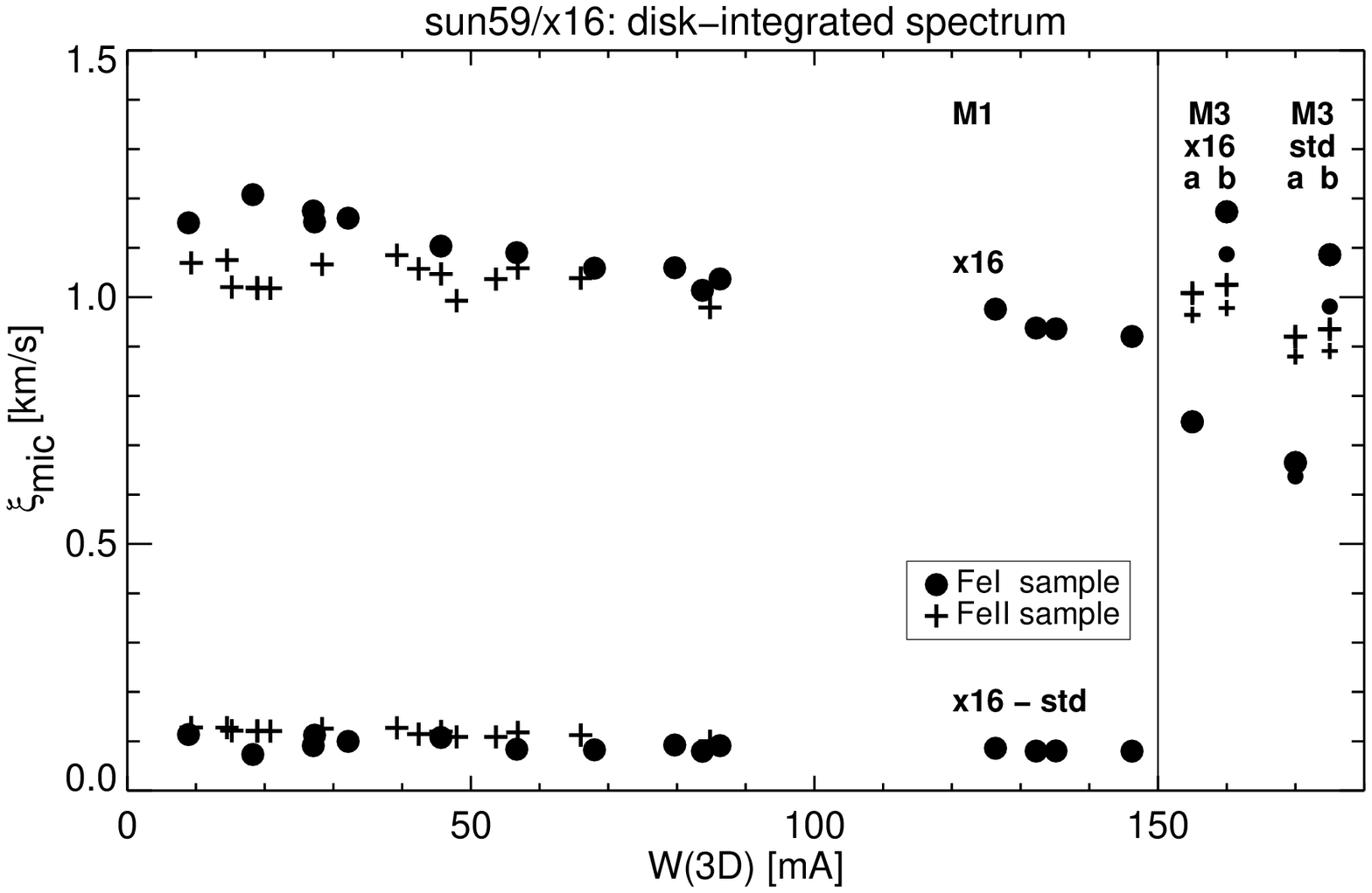}}
\end{center}
\caption{\footnotesize
Same as Fig.\,\ref{f:sun1}, but for one set of 15 real \ion{Fe}{i} lines
with  $504$\,nm $< \lambda < 698$\,nm, $1.5$\,eV $< E_i < 4.6$\,eV (dots),
and one set of 15 real \ion{Fe}{ii} lines with 
$457$\,nm $< \lambda < 772$\,nm, $2.8$\,eV $< E_i < 3.9$\,eV (plus signs).
}
\label{f:sun3}
\end{figure*}
The fact that the derived microturbulence parameter depends on the
considered ion, excitation potential, and line strength, shows that 
a constant microturbulence is not fully appropriate for representing the 
3D hydrodynamical velocity field.
In Fig.\,\ref{f:sun2} we have plotted the microturbulence results
(from disk-integrated spectra only) as a function of the line's
height of formation, to see whether the concept of a depth-dependent
microturbulence might yield a more consistent picture. It is clear that 
\ximic\ is not simply a function of optical depth; it depends also on other
properties of the line, in particular on the vertical extent of the line 
formation region. For this reason, a depth-dependent microturbulence model 
seems not very appealing.

\begin{figure*}[htb]
\begin{center}
\mbox{\includegraphics[bb=40 28 570 370,width=0.75\textwidth,clip=true]
{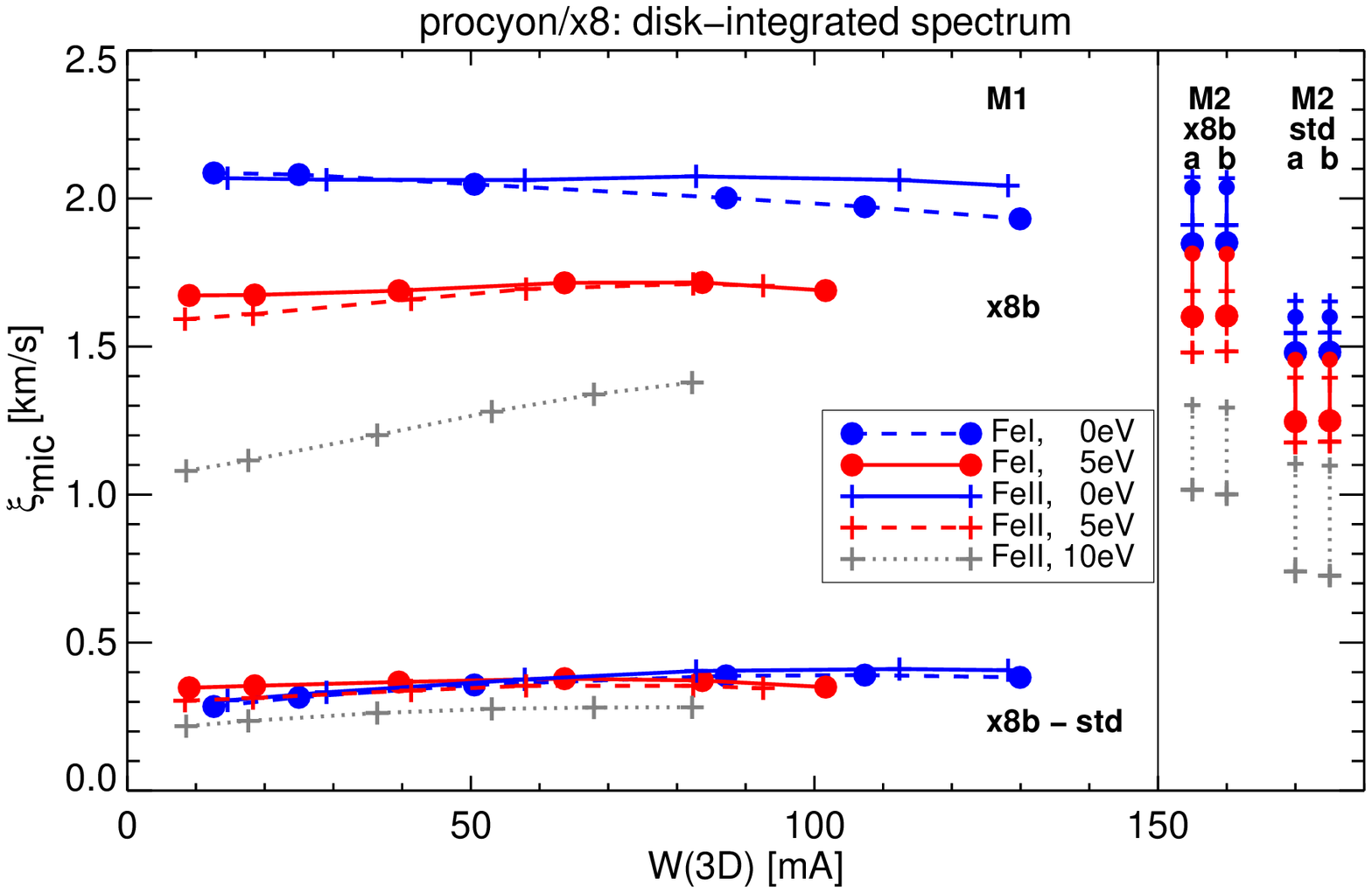}}
\mbox{\includegraphics[bb=40 28 570 370,width=0.75\textwidth,clip=true]
{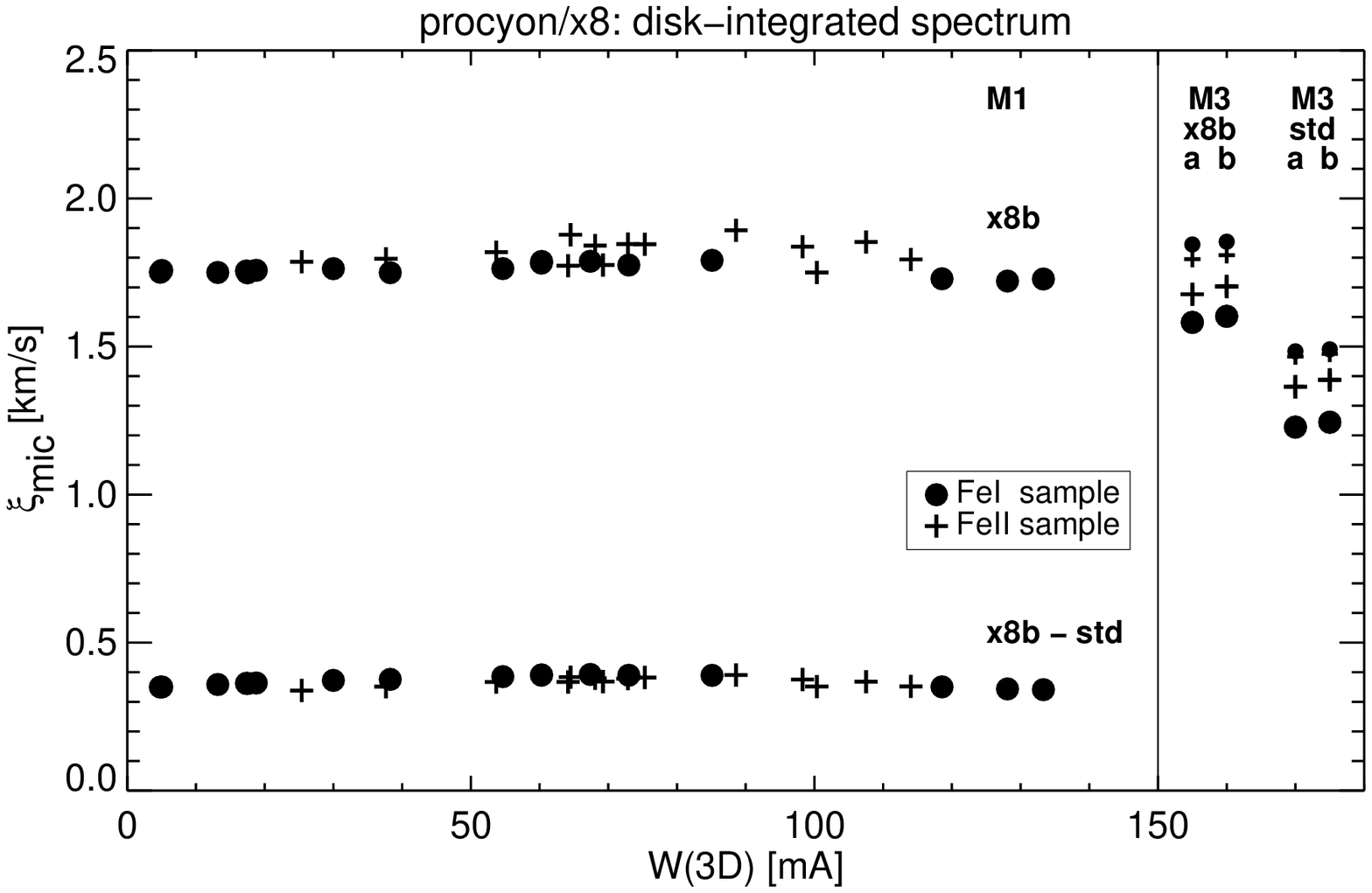}}
\end{center}
\caption{\footnotesize
Determination of the microturbulence parameter \ximic\ for two different
Procyon models (t65g40mm00std and t65g40mm00x8b) from disk-integrated
synthetic (flux) spectra according to methods M1, M2, and M3.
\emph{Upper panel:} \ximic\ obtained from $5$ sets of $6$ fictitious 
iron lines each (see legend) at $\lambda\,550$~nm. \emph{Lower panel:} 
\ximic\ obtained from two sets of real lines ($18$ \ion{Fe}{i} and $14$
\ion{Fe}{ii} lines, respectively). 
The left part of each panel shows the results of M1, where \ximic(t65g40mm00x8b)
is shown together with the difference \ximic(t65g40mm00x8b) - \ximic(t65g40mm00std).
The right part of the upper panel displays the results of M2 for both Procyon
models, in the same representation as in Fig.\,\ref{f:sun1}
(smaller symbols: $\langle$3D$\rangle$, larger symbols: LHD). Similarly, the 
results obtained with M3 are shown in the right part of the lower panel.
}
\label{f:pcy1}
\end{figure*}

\begin{figure*}[htb]
\begin{center}
\mbox{\includegraphics[bb=40 28 570 370,width=0.75\textwidth,clip=true]
{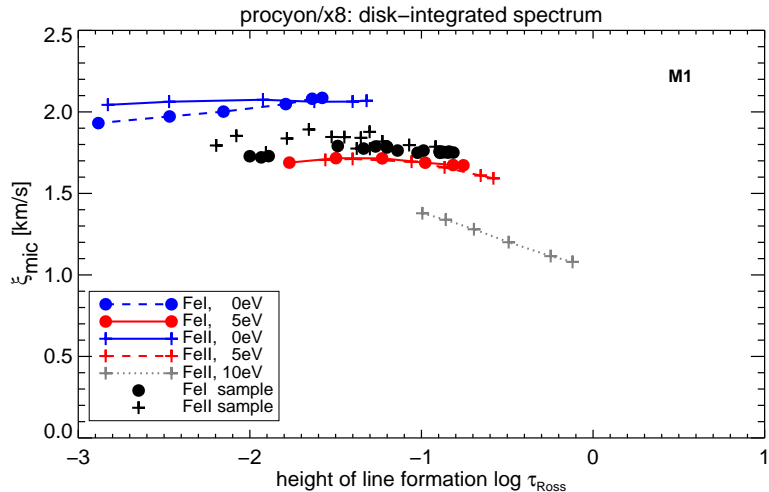}}
\end{center}
\caption{\footnotesize
Microturbulence parameter \ximic\ derived from high-resolution Procyon model 
t65g40mm00x8b according to method M1. The results are the same as shown in
in Fig.\,\ref{f:pcy1}, but are here plotted over the line's height of 
formation (cf.\ Fig.\,\ref{f:sun2}).
}
\label{f:pcy2}
\end{figure*}

\subsubsection{Method 2}
In general, M2a and M2b produce very similar results for a given 1D reference 
atmosphere (see Fig.\,\ref{f:sun1}, right sub-panels). This means that the 
`zero slope' and the `minimum dispersion' conditions are satisfied almost
simultaneously, which is not surprising in the present case of a homogeneous 
sample of lines with identical $\lambda$ and $E_i$. However, we have no
simple explanation for the fact that M2b always yields a slightly higher
\ximic\ than M2a. For the 1D reference 
atmosphere, we have used the averaged 3D model, $\langle$3D$\rangle$, and the
Holweger-M\"uller atmosphere \citep{hmsunmod}, HM. The difference
\mbox{$|\,$\ximic($\langle$3D$\rangle$) - \ximic(HM)$\,|$} derived with M2ab 
is significant but small ($\la 0.1$~km/s).

For \ion{Fe}{i}, the microturbulence derived from M2 agrees closely with
the minimum \ximic\ values obtained from M1 for the respective set of lines,
i.e.\ with \ximic\ obtained from M1 for the stronger lines.  
This is also true for the low-excitation \ion{Fe}{ii} lines 
($E_i=0$~eV). However, M2 gives significantly lower values of \ximic\ 
than M1 for the high-excitation \ion{Fe}{ii} lines. This is particularly clear
for \ion{Fe}{ii}, $E_i$=$10$~eV, at disk center: While M1 gives 
\ximic\ $\approx 0.55$~km/s, consistently for weak and strong lines, M2b gives 
\ximic($\langle$3D$\rangle$) = $0.20$~km/s, and a negative \ximic(HM);
M2a fails to find a positive solution for \ximic\ with any of the 
1D reference atmospheres. In other words, M2a requires a negative 
microturbulence in both of the 1D reference atmospheres to obtain a uniform 
abundance from the \ion{Fe}{ii}, $E_i=10$~eV lines of different strength.
This behavior is explained by deviations between the 1D and 3D thermal 
structures that lead to line strength dependent abundance corrections, which
can only be compensated by a negative value of \ximic. Note that this bias 
is not present in M1, which relies on a comparison of two models with 
identical thermal structure.


Adopting the microturbulence parameter obtained from \ion{Fe}{i} lines
of intermediate excitation potential ensures that the abundance derived
from these lines will not depend systematically on line strength. With this
choice of \ximic, however, the high-excitation \ion{Fe}{ii} lines will show 
a systematic line strength dependence, in the sense that stronger lines 
indicate lower abundances.

Finally, we note that M2 shows the same dependence on the numerical
resolution of the 3D model atmospheres as M1.

\subsubsection{Method 3}
In addition to the fictitious lines discussed above, we have also employed 
two samples of \emph{real} iron lines. Sample~1 comprises $15$ \ion{Fe}{i} 
lines in the wavelength range $504$\,nm $< \lambda < 698$\,nm, with excitation 
potential $1.5$\,eV $< E_i < 4.6$\,eV, sample 2 consists of $15$ \ion{Fe}{ii} 
lines with $457$\,nm $< \lambda < 772$\,nm, $2.8$\,eV $< E_i < 3.9$\,eV
\citep[see][Table.\,2]{caffau2011}.
As expected, the microturbulence determined from these lines with M1 fall 
within the results obtained from the fictitious lines 
(see Fig.\,\ref{f:sun3}). The wavelength dependence of \ximic\ is 
of minor importance in the considered wavelength range.

For both the \ion{Fe}{i} and the  \ion{Fe}{ii} sample, the results of M3b 
and M1 are in reasonable agreement. Also, M3a and M3b give very similar
answers in the case of the \ion{Fe}{ii} sample. For the \ion{Fe}{i} sample,
however, M3a indicates a much lower \ximic\ value than M3b 
(see Fig.\,\ref{f:sun3}, right sub-panels). Obviously, M3a is more 
susceptible to the detailed properties of an inhomogeneous sample of spectral
lines. In particular, the resulting \ximic\ can easily be biased by a 
correlation between excitation potential and line strength. For this reason, 
we prefer in general M3b over M3a.

Like M1 and M2, we find that also M3 indicates systematically higher 
\ximic\ values, by about $0.1$~km/s, for the high-resolution 3D model 
atmosphere of the Sun (sun59x16).

\subsection{\ximic\ for Procyon}
Figure \ref{f:pcy1} shows the determination of the 
microturbulence parameter from two different 3D model atmospheres 
representative of Procyon (t65g40mm00std and t65g40mm00x8b; 
see Tab.\,\ref{tab:3dmodels}), according to methods M1, M2, and M3 
applied to the disk-integrated synthetic flux spectra.
We considered the same $5$ sets of fictitious 
iron lines as for the Sun, but with rescaled $\log gf$-values to adjust
the line strengths such that they fall in the range between $5$ and 
$150$~m\AA\ ($\lambda = 550$~nm).


\begin{figure*}[htb]
\begin{center}
\mbox{\includegraphics[bb=40 28 570 370,width=0.75\textwidth,clip=true]
{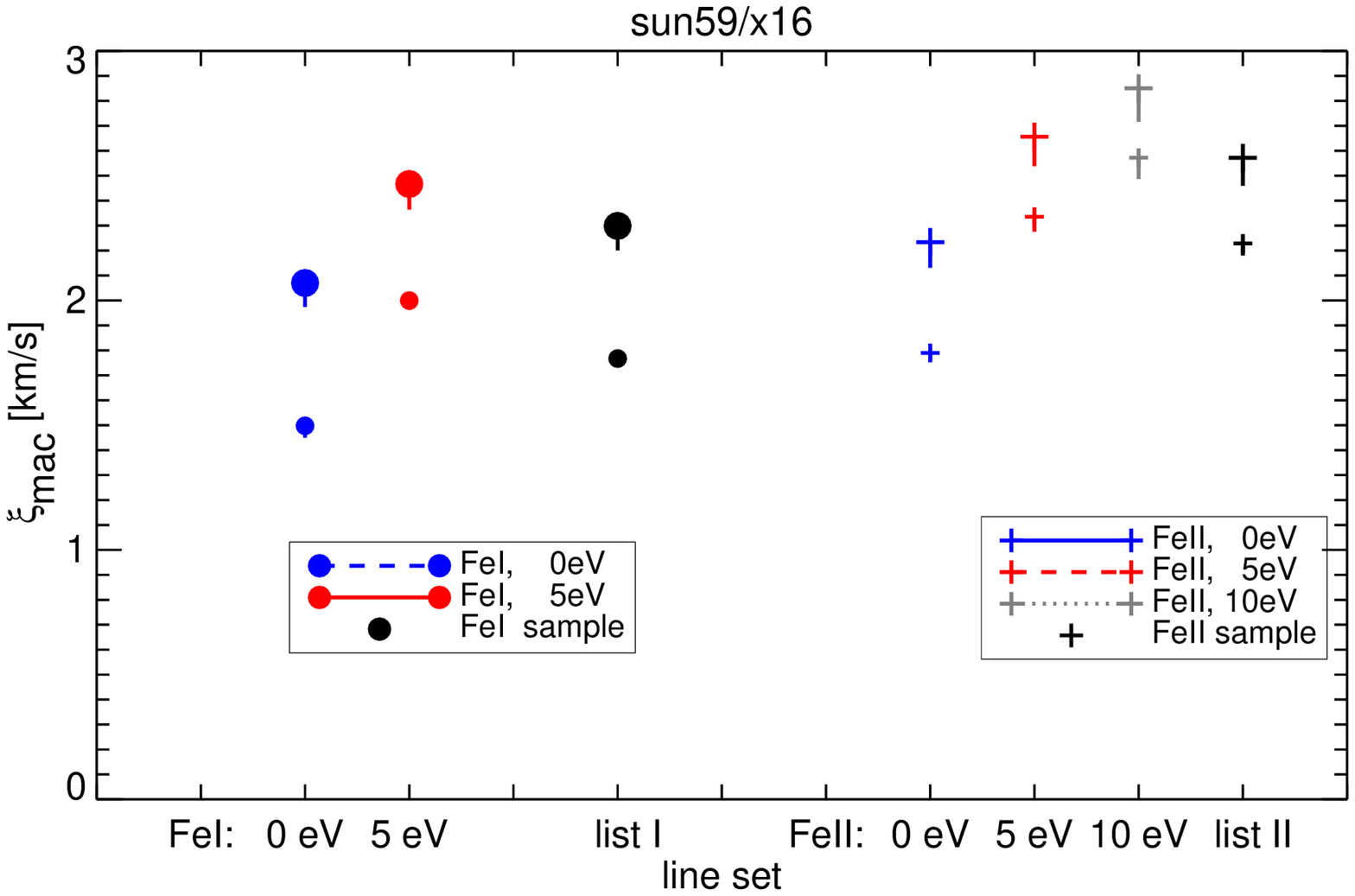}} \hspace*{2mm}
\mbox{\includegraphics[bb=40 28 570 370,width=0.75\textwidth,clip=true]
{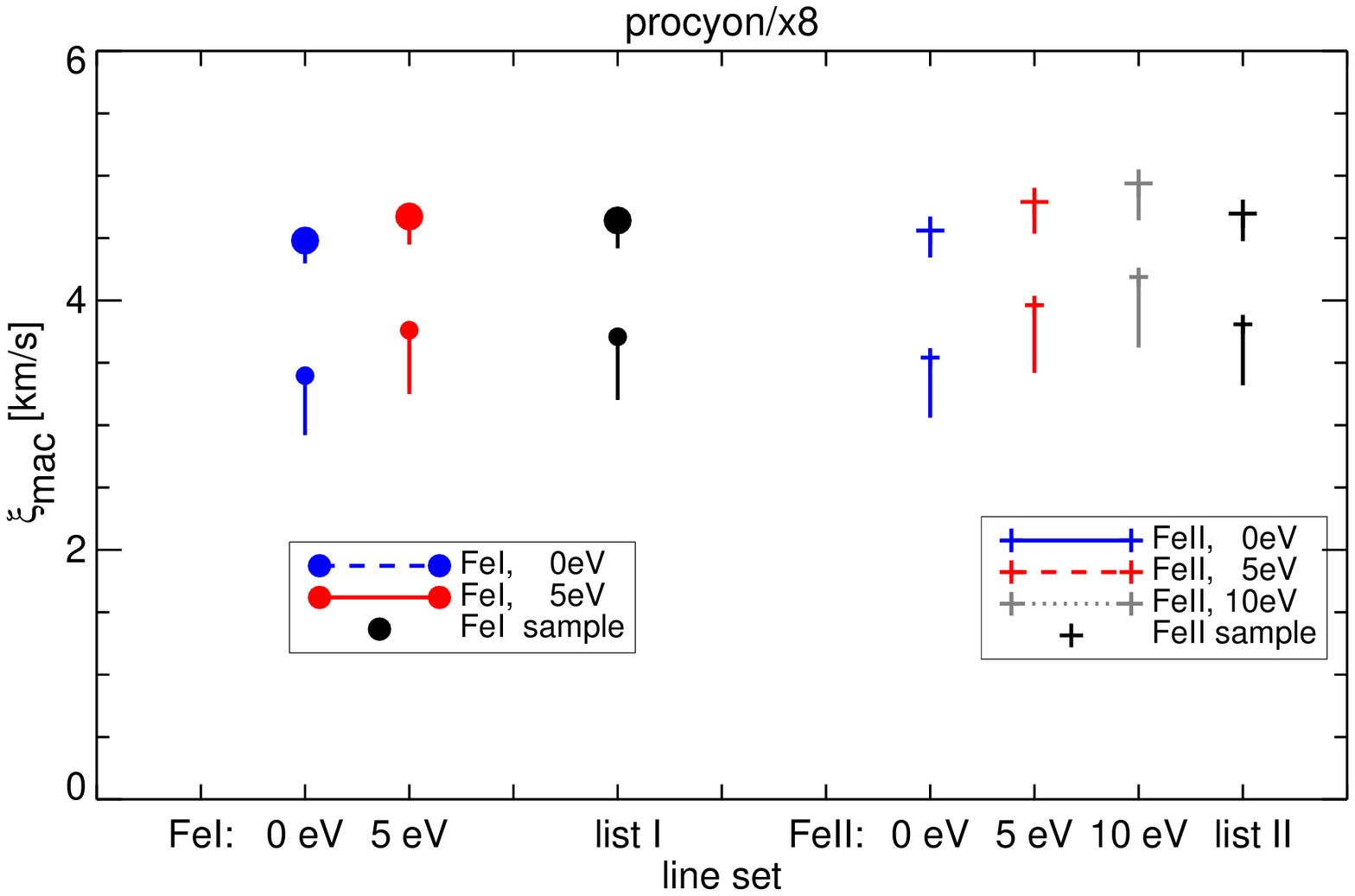}}
\end{center}
\caption{\footnotesize
Summary of macroturbulence determinations for the Sun (upper) and Procyon
(lower) based on Method~1. Each symbol marks the average \ximac\ value 
obtained with the high-resolution 3D model from the same set of lines 
that was used for the derivation of \ximic, as indicted in the legend and
on the abscissa: two sets of fictitious and one sample of real \ion{Fe}{i} 
lines (filled dots) and three sets of fictitious and one sample of real
\ion{Fe}{ii} lines (plus signs). Vertical lines connect each symbol to the
corresponding result obtained with the standard (lower resolution) 3D
model atmosphere. Smaller and larger symbols refer to disk-center and
full-disk synthetic spectra, respectively.
}
\label{f:macro}
\end{figure*}

{\bf Method 1} shows a clear dependence of \ximic\ on excitation potential,
in the sense that high-excitation lines indicate a significantly lower
microturbulence than low-excitation lines. This is in agreement with the 
results obtained for the Sun, and seems to indicate that \ximic\ increases
with height. At the same time, the microturbulence derived with M1 for 
given $E_i$ is essentially independent of line strength (except of the extreme
case of the \ion{Fe}{ii} lines with $E_i$=$10$~eV). This result appears to 
contradict the picture of a height-dependent microturbulence. Figure
\ref{f:pcy2} illustrates the complex situation, which obviously cannot
be described by a simple depth-dependence of \ximic.

The results of {\bf method 2} are fully consistent with those of M1.
This is not surprising in view of the fact that \ximic\ derived with
M1 is almost independent of line strength for a given set of lines with 
constant $E_i$. As before, M2a and M2b agree closely. The choice of the 
1D reference atmosphere has a significant influence on the resulting \ximic: 
using the LHD model instead of the averaged $\langle$3D$\rangle$ model 
reduces the microturbulence by about $0.2$~km/s.

In the context of {\bf Method 3}, we have used one sample of $18$ real 
\ion{Fe}{i} lines in the wavelength range $506$\,nm $< \lambda < 562$\,nm, 
with excitation potential $4.2$\,eV $< E_i < 4.5$\,eV, and one sample of 
$14$ \ion{Fe}{ii} lines with $449$\,nm $< \lambda < 646$\,nm, 
$2.8$\,eV $< E_i < 3.9$\,eV. These two sets of lines have only 
$2$ \ion{Fe}{ii} lines in common with the samples of real iron lines used 
for the solar case. Since the range of $\lambda$ and $E_i$ is rather narrow,
the microturbulence determined from these lines with M3 is consistent with
the very uniform results obtained with M1 (see Fig.\,\ref{f:pcy1}). M3a and
M3b give essentially the same answer. Again, the \ximic\ values obtained with
the LHD model are lower than those obtained with the $\langle$3D$\rangle$ model.

As for the solar case, we note that the predicted microturbulence values 
depend clearly on the numerical resolution of the 3D model atmospheres used 
for the calculation of the synthetic line profiles.
Increasing the spatial resolution by a factor of $2$ in each of the three
spatial directions, keeping everything else unchanged (t65g40mm00std 
$\rightarrow$ t65g40mm00x8a), leads to an increase of \ximic\ by roughly 
$0.2$~km/s for all type of lines. Reducing in addition the explicit 
turbulent viscosity in the hydrodynamical simulations by a factor $2$ 
(t65g40mm00x8a $\rightarrow$ t65g40mm00x8b) has a distinct impact as well,  
further increasing the spectroscopic microturbulence by $\approx$$0.15$~km/s.
While it is clear that the `standard' Procyon model (t65g40mm00std) 
underestimates \ximic\ significantly (more severely than the standard solar
model underestimates the solar \ximic) it remains unclear which 
microturbulence would be obtained in the limiting case of infinite numerical
resolution.

\begin{table*}[htb]
\caption{Empirical values of $\xi_{\rm mic}$ and $\xi_{\rm mac}$ from the 
literature, compared with the theoretical results derived in this work from 
different 3D CO5BOLD model atmospheres for the Sun and Procyon 
(average of results from \ion{Fe}{i}, $E_i$\,$=$\, $0$ and $5$~eV, obtained
with Method 2ab).}
\label{tab2}
\fontsize{8}{10}\selectfont
\begin{center}
\begin{tabular}{lcccccc}
\noalign{\smallskip}\hline\noalign{\smallskip}
  Atmosphere / & \multicolumn{2}{c}{$\xi_{\rm mic}$ [km/s]} & 
                 \multicolumn{2}{c}{$\xi_{\rm mac}$ [km/s]} &
                 \multicolumn{2}{c}{$v_{\rm turb}=
                 \sqrt{(\xi_{\rm mic}^2+\xi_{\rm mac}^2)/2}$~~[km/s]}\\
  Model        & disk-center      & full-disk   & 
                 disk-center      & full-disk   &
                 disk-center      & full-disk   \\
\noalign{\smallskip}\hline\noalign{\smallskip}
Sun, observed$^{\,a}$      & $1.00\pm 0.15$  & $1.35\pm 0.15$  
                          & $1.63\pm 0.15$  & $1.90\pm 0.15$ 
                          & $1.35\pm 0.10$  & $1.65\pm 0.10$  \\
3D solar models:          &           &          &        &   \\
sun59std                  & $0.70\pm 0.05$  & $0.95\pm 0.10$  
                          & $1.75\pm 0.25$  & $2.20\pm 0.20$ 
                          & $1.35\pm 0.20$  & $1.70\pm 0.15$  \\
sun59x16                  & $0.80\pm 0.05$  & $1.05\pm 0.10$  
                          & $1.75\pm 0.25$  & $2.30\pm 0.20$ 
                          & $1.40\pm 0.20$  & $1.80\pm 0.15$  \\
\noalign{\smallskip}\hline\noalign{\smallskip}
Procyon, observed$^{\,b}$  &  ---            & $2.10\pm 0.30$  
                          &  ---            & $4.20\pm 0.50$ 
                          &  ---            & $3.30\pm 0.30$  \\
3D Procyon models:        &           &          &        &   \\
t65g40mm00std             & $0.95\pm 0.15$  & $1.40\pm 0.20$ 
                          & $3.10\pm 0.25$  & $4.40\pm 0.10$ 
                          & $2.30\pm 0.20$  & $3.30\pm 0.10$  \\
t65g40mm00x8b             & $1.45\pm 0.20$  & $1.85\pm 0.25$ 
                          & $3.60\pm 0.25$  & $4.60\pm 0.10$ 
                          & $2.75\pm 0.20$  & $3.50\pm 0.10$  \\
\noalign{\smallskip}\hline\noalign{\smallskip}
\end{tabular}
\end{center}
Notes: a: \citet{Holweger+al78}, full-disk values interpolated: 
$\xi^2$(full-disk)=$\xi^2$(disk-center)/2 + $\xi^2$(limb)/2;\\ 
b: \citet{Steffen85}
\end{table*}

\subsection{\ximac\ for the Sun and Procyon}
We have derived the macroturbulence parameter \ximac\ from the
3D hydrodynamical model atmospheres using M1, as described in 
Sect.\,\ref{s:macro}. The results for the Sun and Procyon are summarized
in Fig.\,\ref{f:macro}. Ignoring the \ion{Fe}{ii}, $E_i$=$10$~eV lines,
we find for the high-resolution solar model: \ximac\ = $1.9\pm 0.4$~km/s 
(disk-center) and \ximac\ = $2.4\pm 0.3$~ km/s (integrated disk), respectively.
The standard (lower resolution) solar model gives only slightly lower
values. For Procyon, we find a macroturbulence that is roughly twice as 
large as for the Sun. The high-resolution model gives:
\ximac\ = $3.7\pm 0.3$~km/s (disk-center) and 
\ximac\ = $4.7\pm 0.2$~ km/s (integrated disk), respectively. 
The disk-center (integrated-disk) values are lower by $0.5$~km/s ($0.3$~km/s)
when the standard 3D Procyon model with lower spatial resolution and higher
viscosity is used  to derive \ximac.

\subsection{Comparison with observation}
In Table\,\ref{tab2}, we compare empirical micro- and macroturbulence 
determinations from the literature with the theoretical predictions of 
our hydrodynamical model atmospheres for the Sun and Procyon presented
in this work. Somewhat arbitrarily, we consider here only the theoretical 
results obtained with Method 2ab from the fictitious \ion{Fe}{i} lines.

The present investigation confirms the previous preliminary analysis 
by \cite{Steffen+al09} (their Table\,1), suggesting that the
theoretical predictions of $\xi_{\rm mic}$ fall significantly below
the classical empirical estimates, for both solar intensity and flux
spectra, and even more clearly for Procyon. The high-resolution models
(sun59x16 and t65g40mm00x8b) come closer to the empirical \ximic\ values,
but still appear to be too low.

One has to keep in mind, however, that the empirical microturbulence was
not determined in exactly the same way as in the theoretical approach.
For a more reliable quantification of the low-\ximic\ problem, we shall
derive the empirical microturbulence from observed spectra with Method~3, 
with exactly the same set of spectral lines (with well known 
$\log gf$ values) as adopted for the derivation of the theoretical \ximic\ 
values from the synthetic 3D spectra.

The macroturbulence values derived from the CO5BOLD models are somewhat 
larger than deduced from observations, such that the \emph{total} non-thermal
rms velocity $v_{\rm turb}$ (columns (6) and (7) of Tab.\,\ref{tab2}),
and hence the total line broadening, appears to be  very similar in 
simulations and observations. But even this quantity is not entirely 
independent of the numerical resolution of the 3D model atmospheres.

\section{\ximic\ across the HRD}
\begin{figure*}[htb]
\mbox{\includegraphics[bb=28 410 580 740,width=0.98\textwidth,clip=true]
{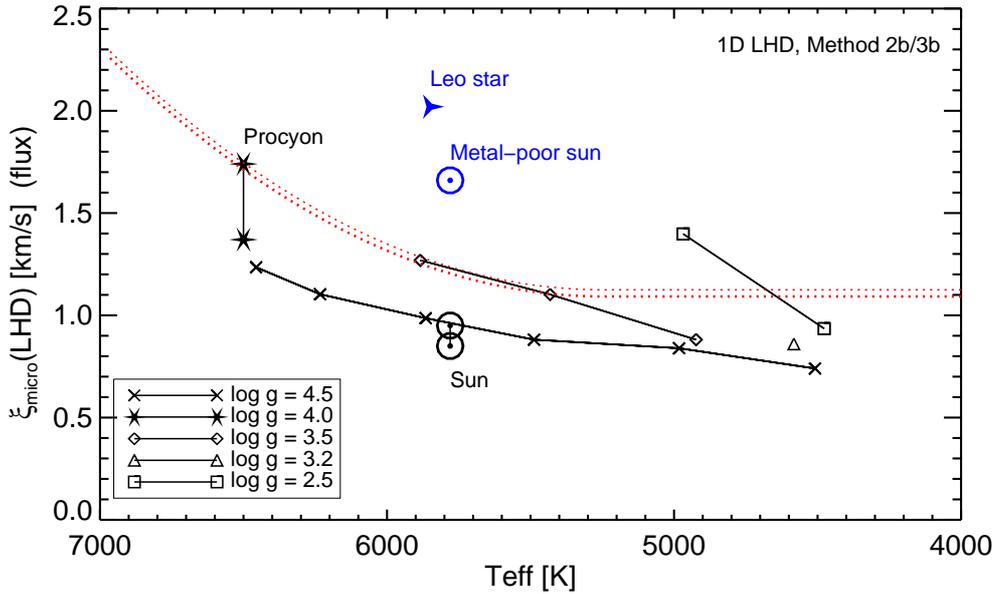}} \hspace*{2mm}
\caption{\footnotesize
Variation of the microturbulence parameter \ximic\ with \teff, \logg, and
\moh, as derived from the disk-integrated synthetic (flux) spectra of a 
number of 3D model atmospheres and the corresponding 1D LHD models 
(see Table\,\ref{tab:xiresults} for more details). Different symbols
indicate \emph{Main Sequence} (crosses), \emph{Subgiants} (diamonds),
and \emph{Giants} (squares). The two dotted curves 
represent an empirical microturbulence parameterization for main sequence
stars (see text; lower and upper curve: \logg=4.5 and 4.0, respectively).
}
\label{f:hrd}
\end{figure*}

In addition to the Sun and Procyon, further 3D hydrodynamical model
atmospheres, taken from the CIFIST 3D model atmosphere grid
\citep{Ludwig+al09} and listed in Table\,\ref{tab:3dmodels}, have been 
analyzed to get a first idea of how the predicted microturbulence 
varies across the Hertzsprung-Russell diagram. The results are given in
Table\,\ref{tab:xiresults} and Fig.\,\ref{f:hrd}. For the $12$ models 
representing the \emph{Main Sequence}, \emph{Subgiants}, and \emph{Giants}, 
we have obtained \ximic\ from a list for $26$ real \ion{Fe}{i} lines 
with $E_i>1.6$~eV, evaluated with M3b. For reference, we have added 
the Sun, Procyon, and the two metal-poor models from a different set of
calculations, where we have instead used a set of $5$ fictitious \ion{Fe}{i} 
lines with $E_i=3.0$~eV ($\lambda\,550$~nm) to derive \ximic\ according 
to M2b. In all cases, we have analyzed the disk-averaged flux spectra of 
the 3D model in comparison with two different 1D atmospheres, namely the 
LHD model with the same stellar parameters as the 3D model, and the average 
$\langle$3D$\rangle$ model. 

\begin{table}
\caption{Microturbulence derived from the disk-integrated (flux) spectra of 
a number of 3D model atmospheres probing a limited region of solar-like 
stars in the HRD.}
\label{tab:xiresults}
\begin{center}
\begin{tabular}{lccc}
\noalign{\smallskip}\hline\noalign{\smallskip}
3D Model & Meth.\  & \ximic\ & $\alpha_{\rm MLT}$   \\
         &         & [km/s ] &                    \\
\noalign{\smallskip}\hline\noalign{\smallskip}
Sun: \\
sun59std  &  M2b & 0.85 ~(0.94) & 0.5  \\
sun59x16  &  M2b & 0.95 ~(1.03) & 0.5  \\
\noalign{\smallskip}\hline\noalign{\smallskip}
Procyon: \\
t65g40mm00std  &  M2b & 1.37 ~(1.53) & 0.5  \\
t65g40mm00x8b  &  M2b & 1.74 ~(1.94) & 0.5  \\
\noalign{\smallskip}\hline\noalign{\smallskip}
Main sequence: \\
t45g45mm00n01  &  M3b & 0.74 ~(0.48) & 0.5 \\
t50g45mm00n04  &  M3b & 0.84 ~(0.65) & 0.5 \\
t55g45mm00n01  &  M3b & 0.88 ~(0.80) & 0.5 \\
t59g45mm00n01  &  M3b & 0.99 ~(0.97) & 0.5 \\
t63g45mm00n01  &  M3b & 1.10 ~(1.14) & 0.5 \\
t65g45mm00n01  &  M3b & 1.24 ~(1.26) & 0.5 \\
\noalign{\smallskip}\hline\noalign{\smallskip}
Subgiants: \\
t46g32mm00n01  &  M3b & 0.86 ~(0.80) & 0.5 \\
t50g35mm00n01  &  M3b & 0.88 ~(0.85) & 0.5 \\
t55g35mm00n01  &  M3b & 1.10 ~(1.20) & 0.5 \\
t59g35mm00n01  &  M3b & 1.27 ~(1.42) & 0.5 \\
\noalign{\smallskip}\hline\noalign{\smallskip}
Giants: \\
t45g25mm00n01  &  M3b & 0.93 ~(0.93) & 0.5 \\
t50g25mm00n01  &  M3b & 1.40 ~(1.51) & 1.0 \\
\noalign{\smallskip}\hline\noalign{\smallskip}
Metal-poor sun: \\
t57g44mm20n03  &  M2b & 1.66 ~(0.82) & 1.0 \\
\noalign{\smallskip}\hline\noalign{\smallskip}
Leo star: \\
t59g40mm40n02  &  M2b & 2.02 ~(0.98) & 0.5 \\
\noalign{\smallskip}\hline\noalign{\smallskip}
\end{tabular}
\end{center}
Notes: $\alpha_{\rm MLT}$ is the mixing length parameter used for
the calculation of the 1D LHD reference model. Results obtained with 
Method 2b and 3b were obtained with a set of 5 fictitious \ion{Fe}{i} 
lines ($E_i$=$3$~eV, $\lambda\,550$~nm) and a set of 26 real 
\ion{Fe}{i} lines ($E_i>1.6$~eV), respectively. Numbers in parenthesis 
refer to the results obtained when replacing the LHD model with the 
$\langle$3D$\rangle$ model.
\end{table}

Figure \ref{f:hrd} shows that the predicted \ximic\ increases systematically 
towards higher \teff\ and lower \logg. This trend is in agreement with 
empirical evidence. However, we note that quantitatively the 
\ximic\ values predicted from the standard (low resolution, high viscosity)
3D model atmospheres are lower, by roughly $0.3$~km/s, than those obtained 
from the empirical relation recommended by the Gaia-ESO
consortium\footnote{http://great.ast.cam.ac.uk/GESwiki/GesWg/
GesWg11/Microturbulence}. For
main sequence and subgiants, this relation is given by
\begin{eqnarray}
\xi_{\rm mic} &=& 1.15 + 2\,10^{-4}\,X + 3.95\,10^{-7}\,X^2
\nonumber \\
            &-& 0.13\,Y + 0.13\,Y^2 \quad \mathrm{[km/s]}\, , 
\label{eq1}
\end{eqnarray}
with $X \equiv \max\{-250,T_{\rm eff}-5500\}$, $Y\equiv\log g-4$.
It is indicated for \logg=4.5 and 4.0 by the (red) dotted lines in 
Fig.\,\ref{f:hrd}. The high-resolution Sun (sun59x16) lies clearly below 
this empirical relation, while the high-resolution Procyon model 
(t65g40mm00x8b) agrees closely.

Curiously, the two metal-poor models indicate an anomalously high 
microturbulence. This is, however, not a sign of increased turbulence 
in metal-poor stellar atmospheres. As evident from Table\,\ref{tab:xiresults},
this anomaly vanishes when doing the microturbulence determination with the
$\langle$3D$\rangle$ model as the 1D reference. We conclude that the high
microturbulence obtained with the LHD model is related to the fact that the
temperature structures of LHD and $\langle$3D$\rangle$ model deviate strongly,
in the sense that the $\langle$3D$\rangle$ model is much cooler in the upper 
photosphere. As a consequence, matching the 3D equivalent width of a
low-excitation iron line with the LHD model requires a much higher iron 
abundance for strong lines than for weak lines, unless this mismatch is 
compensated by a high microturbulence. 
Observationally, this high-\ximic\ effect can probably not be verified
due to the absence of sufficiently strong iron lines.

\section{Discussion and conclusions}
We have applied different methods to extract the parameters $\xi_{\rm mic}$
and $\xi_{\rm mac}$ from synthetic spectra based on 3D hydrodynamical 
atmospheres of a number of solar-type stars. We find that the different
methods give consistent results. However, the derived turbulence parameters
depend systematically on the properties of the selected spectral lines.
It is thus not obvious how to assign a unique value of \ximic\ to a given 
stellar atmosphere.

For the Sun and Procyon, we have also demonstrated that the numerical
viscosity and spatial resolution of the 3D model atmospheres still has a 
significant influence of the resulting micro- and macroturbulence, in the 
sense that the high-resolution models imply somewhat higher turbulence values.
A preliminary comparison of the \ximic\ values predicted from the 3D 
simulations with the results of empirical studies found in the literature 
reveals that the theoretical predictions are systematically too low
(see Table\,\ref{tab2}). The additional models included in the present 
study confirm this conclusion: a similar microturbulence discrepancy is 
seen for all solar-type main sequence stars (Fig.\,\ref{f:hrd}). 

These findings suggests that the 
velocity field provided by the standard 3D hydrodynamical models is less 
`turbulent' than it is in reality. While this conclusion seems to be in 
conflict with the claims by \citet{A2000}, it is confirmed by 
\cite{2002ApJ...567..544A}, who find from their 3D Procyon model a systematic 
increase of the iron abundance with line strength, both for \ion{Fe}{i} and 
\ion{Fe}{ii} lines (see their Fig.\,17). This translates to a microturbulence 
deficiency of about $0.3$~km/s, and implies a systematic overestimation of 
3D abundances from stronger lines. 

Our present results indicate that the discrepancy between theoretical and
empirical \ximic\ is reduced, but not completely removed, when the latest 
3D high-resolution model atmospheres are utilized.
For a more reliable quantification of the remaining gap, we shall
not rely on literature values of \ximic, but instead intend to derive the 
empirical microturbulence from observed spectra with Method~3, in exactly 
the same way and with the same set of spectral lines (with well known 
$\log gf$ values) as adopted for the derivation of the theoretical \ximic\ 
values from the synthetic 3D spectra.

Further investigations are necessary to map the microturbulence problem
across the Hertzsprung-Russell diagram, and to find an appropriate recipe
to make the best use of the 3D models. A new generation of high-resolution, 
low viscosity 3D model atmospheres is certainly welcome in this respect. 
But it might still be necessary to introduce some (resolution-dependent) 
3D microturbulence component in addition to the large-scale hydrodynamical
velocity field of the 3D simulations, just for the purpose of an accurate 
representation of the non-thermal Doppler broadening of stronger spectral 
lines. Conceivably, the required small-scale velocity enhancement can be
predicted by an appropriate turbulence model.

\begin{acknowledgements}
EC and HGL acknowledge financial support by the Sonderforschungsbereich
SFB\,881 `The Milky Way System' (subproject A4) of the German Research
Foundation (DFG).
\end{acknowledgements}

\bibliographystyle{aa}

\begin{thebibliography}{}

\bibitem[Allende Prieto et al.(2002)]{2002ApJ...567..544A} Allende Prieto, 
C., Asplund, M., Garc{\'{\i}}a L{\'o}pez, R.J., Lambert, D.L.\ 2002, 
\apj, 567, 544 

\bibitem[Asplund et al.(2000)]{A2000} Asplund, M., Nordlund, {\AA}., 
Trampedach, R., Allende Prieto, C., \& Stein, R.~F.\ 2000, \aap, 359, 729 

\bibitem[Caffau et al.(2011)]{caffau2011} Caffau, E., Ludwig, 
H.-G., Steffen, M., Freytag, B., \& Bonifacio, P.\ 2011, \solphys, 268, 255 

\bibitem[Freytag et al. (2012)]{Freytag+al12}
Freytag, B., Steffen, M., Ludwig, H.-G., et al.\ 2012, 
Journal of Computational Physics, 231, 919

\bibitem[Holweger \& M\"uller(1974)]{hmsunmod} Holweger, H., \& 
M\"uller, E.A.\ 1974, \solphys, 39, 19 

\bibitem[\protect\citeauthoryear{Holweger et al.}{1978}]{Holweger+al78} 
Holweger, H., Gehlsen, M., \& Ruland, F. 1978, 
\aap, 70, 537

\bibitem[Ludwig et al. (2009)]{Ludwig+al09}
Ludwig, H.-G., Caffau, E., Steffen, M., et al.\ 2009, \memsai, 80, 711 

\bibitem[Markwardt(2009)]{Markwardt2009} Markwardt, C.B.\ 2009, 
Astronomical Data Analysis Software and Systems XVIII, 411, 251 

\bibitem[\protect\citeauthoryear{Steffen}{1985}]{Steffen85}
Steffen, M. 1985, \aaps, 59, 403

\bibitem[Steffen et al.(2009)]{Steffen+al09}
Steffen, M., Ludwig, H.-G., \& Caffau, E.\ 2009, \memsai, 80, 731 

\end{thebibliography}

\begin{appendix}
\section{Direct derivation of \ximic\ and \ximac\ 
         from the hydrodynamical velocity field}

A possible approach to deriving the disk-center ($\mu=1$) values  
of $\xi_{\rm mic}$ and $\xi_{\rm mac}$ directly from the 3D hydrodynamical 
velocity field is as follows:

First, a weighting function $w(\tau)$ is defined on a standard optical
depth scale, e.g.\ $\tau=\tau_{\rm Ross}$, describing the
contribution of the different photospheric layers to the formation
of a typical line. Presumably, the weighting function is closely related
to the line depression contribution function. It is normalized as
\begin{equation}
\int_0^\infty w(\tau) \; \mathrm{d} \tau = 1 \, .
\end{equation}
Next, we compute the depth-weighted first and second moments of the 
vertical velocity $u_z$ at each horizontal position $(x,y)$ for all 
selected snapshots $t$:
\begin{equation}
\overline{u_z}(x,y,t) = \int_0^\infty u_z(x,y,\tau,t)\,w(\tau) \; \mathrm{d} \tau\, ,
\end{equation}
and
\begin{equation}
\overline{u_z^2}(x,y,t) = \int_0^\infty u_z^2(x,y,\tau,t)\,w(\tau) \; \mathrm{d} \tau\, .
\end{equation}
The macroturbulence parameter is then computed as the variance of the
local line shift $\overline{u_z}(x,y,t)$ over the stellar surface,
\begin{equation}
\xi_{\rm mac}^2 = 2\, \left(\,\left\langle \overline{u_z}^2(x,y,t) \right\rangle\, - 
                    \left\langle \strut \overline{u_z}(x,y,t) \right\rangle^2\,\right)\,,
\end{equation}
and the microturbulence parameter as the ($x,y,t$)-average of the local
line-of-sight velocity dispersion,
\begin{equation}
\xi_{\rm mic}^2 = 2\, \left(\,\left\langle \overline{u_z^2}(x,y,t) \right\rangle\, - 
                         \left\langle \overline{u_z}^2(x,y,t) \right\rangle\,\right)\,,
\end{equation}
where $\langle.\rangle$ denotes horizontal averaging over $(x,y)$ and 
temporal  averaging over $t$. The total turbulent velocity is then
\begin{eqnarray}
v_{\rm turb}^2&=& \left(\xi_{\rm mic}^2 + \xi_{\rm mac}^2\right)/2 \nonumber \\
             &=& \,\left\langle \overline{u_z^2}(x,y,t) \right\rangle\, - 
 \left\langle \strut \overline{u_z}(x,y,t) \right\rangle^2\, .
\end{eqnarray}
Taking into account the horizontal components of the hydrodynamical
velocity field, $u_x$ and $u_y$, the procedure can be generalized to 
evaluate $\xi_{\rm mic}$  and $\xi_{\rm  mac}$ values also for the 
disk-averaged spectrum: 
\begin{eqnarray}
\xi_{\rm mac}^2 &=& \frac{1}{2}\, \left(\,\left\langle \overline{u_x}^2(x,y,t) \right\rangle\, - 
                      \left\langle \strut \overline{u_x}(x,y,t)
                      \right\rangle^2\,\right) \nonumber \\
              &+& \frac{1}{2}\, \left(\,\left\langle \overline{u_y}^2(x,y,t) \right\rangle\, - 
                      \left\langle \strut \overline{u_y}(x,y,t)
                      \right\rangle^2\,\right) \nonumber \\
              &+&     \left(\,\left\langle \overline{u_z}^2(x,y,t) \right\rangle\, - 
                      \left\langle \strut \overline{u_z}(x,y,t)
                      \right\rangle^2\,\right) \, ,
\end{eqnarray}
\begin{eqnarray}
\xi_{\rm mic}^2 &=& \frac{1}{2}\, \left(\,\left\langle \overline{u_x^2}(x,y,t) \right\rangle\, - 
                      \left\langle \strut \overline{u_x}^2(x,y,t)
                      \right\rangle\,\right) \nonumber \\
              &+& \frac{1}{2}\, \left(\,\left\langle \overline{u_y^2}(x,y,t) \right\rangle\, - 
                      \left\langle \strut \overline{u_y}^2(x,y,t)
                      \right\rangle\,\right) \nonumber \\
              &+&     \left(\,\left\langle \overline{u_z^2}(x,y,t) \right\rangle\, - 
                      \left\langle \strut \overline{u_z}^2(x,y,t)
                      \right\rangle\,\right) \, ,
\end{eqnarray}
\begin{eqnarray}
v_{\rm turb}^2 &=& \frac{1}{4}\, \left(\,\left\langle \overline{u_x^2}(x,y,t) \right\rangle\, - 
                      \left\langle \strut \overline{u_x}(x,y,t)
                      \right\rangle^2\,\right) \nonumber \\
            &+& \frac{1}{4}\, \left(\,\left\langle \overline{u_y^2}(x,y,t) \right\rangle\, - 
                      \left\langle \strut \overline{u_y}(x,y,t)
                      \right\rangle^2\,\right) \nonumber \\
            &+& \frac{1}{2}\, \left(\,\left\langle \overline{u_z^2}(x,y,t) \right\rangle\, - 
                      \left\langle \strut \overline{u_z}(x,y,t)
                      \right\rangle^2\,\right) \, .
\end{eqnarray}

First experiments have shown that the \ximic\ and \ximac\ values obtained
with this method fall in the same range as those derived with the
spectroscopic approach. However, the results depend sensitively on the choice
of the weighting function $w$. Further thoughts are necessary to
work out an appropriate definition of $w$. It might turn out that $w$ 
does not depend on optical depth only.

\end{appendix}

\end{document}